\documentclass[12pt]{article}
\usepackage{ amsmath, amsthm, amssymb } 
\usepackage{adjustbox}

\usepackage{url}  
\usepackage{color} 
\usepackage{cite}

\newcommand{\be} { \begin{equation} } 
\newcommand{\ee} { \end{equation} } 
\newcommand{\I}{\mathcal{I}}

\newcommand{\fin}{{\rm fin}}
\newcommand{\Li}{{\rm Li}}
\newcommand{\uv}{\rm UV}
 
\newcommand{\p}{p\hspace{-0.9ex}/}

\newcommand{\un}{{\rm un}}
\begin{document} 
\allowdisplaybreaks

\setlength{\unitlength}{1.3cm} 
\begin{titlepage}
\vspace*{-1cm}
\begin{flushright}
TTP17-002
\end{flushright}                                
\vskip 3.5cm
\begin{center}
\boldmath

{\Large\bf Two-loop amplitudes for $q g \to H q$ and $q \bar{q} \to H g$ 
mediated by a nearly massless quark \\[3mm] }
\unboldmath
\vskip 1.cm
{\large Kirill Melnikov}$^{a,}$
\footnote{{\tt e-mail: kirill.melnikov@kit.edu}}
{\large , Lorenzo Tancredi}$^{a,}$
\footnote{{\tt e-mail: lorenzo.tancredi@kit.edu}} 
{\large , Christopher Wever}$^{a,b,}$
\footnote{{\tt e-mail: christopher.wever@kit.edu}} 
\vskip .7cm
{\it $^a$  Institute for Theoretical Particle Physics, KIT, 76128 Karlsruhe, Germany} \\
{\it $^b$  Institut f\"ur Kernphysik, KIT, 76344 Eggenstein-Leopoldshafen, Germany} 
\end{center}
\vskip 2.6cm

\begin{abstract}
\noindent
We compute the two-loop QCD corrections to $q g \to H q$ and $q \bar{q} \to H g$ amplitudes 
mediated by loops of nearly massless quarks. These amplitudes provide 
the last missing ingredient
required to compute the NLO QCD corrections to the top-bottom
interference contribution to the Higgs boson  transverse momentum distribution at hadron colliders. 

\end{abstract}
\vfill
\end{titlepage}                                                                
\newpage

\section{Introduction} \label{sec:intro} \setcounter{equation}{0} \setcounter{footnote}{0} 
\numberwithin{equation}{section}

A  Standard Model (SM)-like Higgs boson was discovered at the LHC in 2012.  An important 
step in ``fingerprinting'' this particle  is the precise determination of its couplings 
to quarks, leptons and gauge bosons.  Indeed,  
couplings of the SM Higgs boson to other particles should be related to particle masses and 
the exact relation between these particle masses and their couplings to the Higgs can 
be derived  with an  astounding precision in the SM. 
Results from the Run I  of the LHC seem to indicate that  the Higgs  
couplings to SM fermions and vector bosons  are in agreement with the expectations, within 
current   statistical  and systematic errors.

The precision of the Higgs couplings measurements is expected to increase as
 more data from Run II is analyzed; by the end of the high-luminosity phase  
of the LHC it is conceivable that ${\cal O}({\rm few})$-percent  measurements 
of the  Higgs couplings will become available. Motivated by this, 
theoretical predictions for   Higgs boson production cross sections at the LHC 
and for Higgs boson decays  were extended to very high orders in perturbative QCD and 
the SM, resulting in high precision for observables relevant for Higgs phenomenology. 
The  recent highlight of this activity is the  calculation of the inclusive cross section of 
the dominant $gg\rightarrow H$ production process through  N$^3$LO QCD  in the infinite top mass 
limit ~\cite{Anastasiou:2016cez}. However, thanks to increased energy and statistics of  the LHC Run II, 
it is interesting to refine an understanding of more exclusive processes, for example 
production of the Higgs boson in association with a jet. In the $m_t \to \infty$ approximation, the NNLO 
QCD corrections to $pp \to H+j$ and to the Higgs transverse momentum were computed  in
Refs.~\cite{Boughezal:2013uia,Chen:2014gva,Boughezal:2015dra,Boughezal:2015aha}.

While the above computations provide a clear milestone for  applications of perturbative QCD 
to Higgs physics at hadron colliders, there are good reasons to go beyond the $m_t \to \infty$ approximation in theoretical predictions 
for $H+j$ and the Higgs $p_\perp$ spectrum.  Indeed, by measuring the 
Higgs transverse momentum distribution at high $p_\perp$, one can  
probe couplings  of the Higgs boson to hypothetical top partners and  
constrain    models beyond the SM  \cite{Arnesen:2008fb}. A particular  valuable is the 
kinematic region where the Higgs boson is produced with  a large transverse 
momentum~\cite{Harlander:2013oja,Azatov:2013xha,Grojean:2013nya,Banfi:2013yoa}.\footnote{See~\cite{Neumann:2016dny} for further references.} 
A comparison with future high-$p_\perp$ experimental analysis will require 
accurate theoretical predictions for  exclusive Higgs production in association with one or more jets. 

It is also interesting to consider  low and moderate transverse momenta of the produced Higgs bosons 
since this is where the bulk of the 
events is. At values of the Higgs transverse momentum $p_\perp <  m_t $, the top quark loop is, essentially, point-like 
but the bottom quark loop is not.  Naively, the bottom quark loop is expected to be 
 suppressed by a factor of $y_b\, m_b/m_h \sim m_b^2/m_h^2\sim 10^{-3}$ relative to the top quark 
loop but, as it turns out, the reality is more complex. Indeed, it is known that the bottom 
quark loop is enhanced  by a logarithm of the ratio of the Higgs boson 
mass to the $b$-quark mass, $\log^2(m_h^2/m_b^2)\sim {\rm few} \times 10$, so that  
the bottom loop contribution is estimated to change the top loop result by five to ten percent. 

We note that the structure of  logarithms is actually more complicated for observables 
that are less exclusive than the inclusive Higgs production cross section. 
For example, double logarithms of the form $\log^2 (p^2_\perp/m_b^2)$ appear 
in Higgs transverse momentum distribution  for  $p_\perp \gg m_b$ leading to enhanced, 
kinematics-dependent corrections~\cite{Mantler:2012bj,Grazzini:2013mca,Banfi:2013eda,
Bagnaschi:2015bop}. Existing numerical estimates  point towards a few percent 
effects that are caused by top-bottom $p_\perp$-dependent interference contributions. Therefore, since the 
QCD corrections to Higgs boson production in gluon fusion are known to be large and since a few percent 
precision on Higgs production cross sections is a long-term goal of the LHC program, it is 
important to compute the NLO QCD corrections to the bottom loop. 

Another reason to consider  NLO QCD corrections to bottom-quark-mediated contributions  
to Higgs boson production is more theoretical in nature. Indeed, as we already 
mentioned, new double logarithmic terms appear in the perturbative expansion but these 
logarithms are not very well understood so that, for example, their resummation 
can not be currently performed.\footnote{Such corrections 
in the abelian limit and in the high-energy limit were studied to all orders in $\alpha_s$ 
in~Refs.\cite{Melnikov:2016emg,Caola:2016upw}.} 
To this end, explicit NLO calculations of the relevant scattering amplitude provide an important 
``data point'' for future efforts to understand and, perhaps,  resum quark-mass dependent 
 logarithmic corrections in Higgs boson production. 

We have laid the groundwork for such a NLO computation 
in~Ref.~\cite{Melnikov:2016qoc} 
by calculating 
the  $gg \rightarrow Hg$ amplitudes 
mediated by a bottom quark, in the limit $m_b \to 0$.\footnote{A similar approach has been
used to study top-bottom interference effects in single Higgs production in Ref.~\cite{Mueller:2015lrx}.}
 In this paper we compute 
the $m_b\rightarrow 0$ limit 
of the quark-gluon amplitude, relevant for the  remaining partonic channels
$q\bar{q}\rightarrow Hg$ and  $qg\rightarrow Hq$.\footnote{We note that we do not consider 
amplitudes where bottom quark appears as an external particle.}
The calculation of these two-loop amplitudes retaining exact mass dependence 
remains an outstanding task. A first promising step in this direction has been recently taken
with the calculation of the relevant planar master integrals~\cite{Bonciani:2016qxi}.
The results that we present in 
this paper complement those of Ref.~\cite{Melnikov:2016qoc} and, together, 
provide all 
two-loop amplitudes required for the computation of NLO QCD corrections to top-bottom 
interference effects in  the production 
of the Higgs boson in association with a jet, and in the Higgs boson transverse momentum 
distribution. 

The paper is organized as  follows. We explain the notation and 
discuss the Lorentz structure of the relevant amplitudes  in Section \ref{sec:notation}.  
In Sections~\ref{sec:formfactors} and~\ref{sec:UVIR} 
we review the  computation of the form factors and describe their renormalization. We present  
the  results for the helicity amplitudes in Section~\ref{sec:helamp}.
For reasons of convenience, the amplitudes are first computed in the decay kinematics, i.e. for the process  $H \to q\bar{q}g$.  Their  
analytic continuation to the kinematic regions relevant for Higgs boson 
production  is described in Section \ref{sec:cont}. The limit of the helicity amplitudes for the physical scattering 
process $qg \to Hq$, where the initial and final state quark are collinear, is considered in Section~\ref{sec:colimit}.
We conclude in Section~\ref{sec:conclusions}.
Analytic results for the amplitudes in  different kinematic regions are attached as
ancillary files to the arXiv submission of this paper.

\section{Lorentz structure of the scattering amplitude}
\label{sec:notation} \setcounter{equation}{0} 
\numberwithin{equation}{section} 
We consider the process 
\begin{equation}
H(p_4) \to q(p_1) + \bar{q}(p_2) + g(p_3)\,,
\label{eq:Htoqqg}
\end{equation}
mediated by a   bottom quark with the mass $m_b$.
We introduce the usual Mandelstam variables
\be
s = (p_1 + p_2)^2\,, \quad t = (p_1 + p_3)^2\,, \quad u = (p_2 + p_3)^2\,,\qquad\quad
s + t + u = m_h^2,
\ee
where $m_h$ is the Higgs boson mass.
We consider the process in the kinematic  limit where the bottom mass is the smallest
scale in the problem; this implies  that $m_b^2 \ll s \sim  t \sim u \sim m_h^2$.
Following Ref.~\cite{Melnikov:2016qoc},  we introduce the massless ratios
\begin{equation}
x = \frac{s}{m_h^2}\,, \quad y = \frac{t}{m_h^2}\,, \quad z = \frac{u}{m_h^2}\,,
\quad \kappa = - \frac{m_b^2}{m_h^2}\,,
\end{equation}
and consider the amplitude in decay kinematics. This implies 
\begin{equation}
0<y<1\,, \quad 0<z<1\,, \quad 0<x = 1-y-z<1\,, \quad \kappa >0, 
\;\;\; \, m_h^2 > 0\,.
\end{equation}
With these choices, the  only source of imaginary parts in contributing Feynman integrals 
is the positive value of the Higgs mass,
that can be isolated as  an overall prefactor $(- m_h^2 - i\,0)^{-\epsilon}$ per  loop.

We define the partonic scattering amplitude as follows\footnote{We only consider massless quarks 
in the initial and final states.}
\begin{align}
\mathcal{A}(p_1^j, p_2^k, p_3^a) &= i\, T^{a}_{jk}\;
\epsilon^\mu_3(p_3)\, \bar{u}(p_1)\, \mathcal{A}_\mu(s,t,u,m_b)\, v(p_2) \,,
\end{align}
where $j,k$ are the color indices of the quark and the antiquark, respectively,
and $a$ is the color index of the gluon. The amplitude $\mathcal{A}(p_1^j, p_2^k, p_3^a)$ 
should be Lorentz-invariant and transversal 
\begin{equation}
p_3^\mu \;  \bar{u}(p_1)\, \mathcal{A}_\mu(s,t,u,m_b)\, v(p_2) \, = 0.
\label{eq:trans}
\end{equation}
The most general anzats for $\mathcal{A}_\mu$ consistent with these conditions and parity conservation 
involves  two  form factors
\begin{equation}
\mathcal{A}^\mu = F_1\, \left( \p_3\, p_2^\mu - p_2 \cdot p_3  \, \gamma^\mu \right)
+ F_2 \left( \p_3\, p_1^\mu - p_1 \cdot p_3  \, \gamma^{\mu} \right )
= F_1\, \tau_1^\mu + F_2 \, \tau_2^\mu\, . \label{eq:tensor}
\end{equation}
We note that the two tensorial structures in Eq.(\ref{eq:tensor}) 
satisfy the transversality condition Eq.(\ref{eq:trans}) separately. 
The   form factors in Eq.(\ref{eq:tensor}) are Lorentz-scalar functions of the Mandelstam invariants 
$F_j = F_j(s,t,u,m_b)$. 

The unrenormalized form factors $F_j$ can be expanded in the strong coupling constant.
We write 
\begin{equation}
F_j^{\un}(s,t,u,m_b) = \sqrt{\frac{\alpha_0^3}{\pi}} \left[ F_j^{(1),\un} 
+ \left(\frac{\alpha_0}{2 \pi}\right) F_j^{(2),\un}  + \mathcal{O}(\alpha_0^2)\right]\,, \quad j = 1,2,
\label{eq:bareF}
\end{equation}
where $\alpha_0$ is the bare QCD coupling constants. The form factors $F_{1,2}^{(1)}$ 
are known, including the full dependence on the quark mass \cite{Ellis:1987xu,Baur:1989cm}.
Our goal is to compute the NLO QCD contributions to the two form factors in $m_b \to 0$ limit. 

\section{Computation of the form factors}
\label{sec:formfactors} \setcounter{equation}{0} 
\numberwithin{equation}{section}

\begin{figure}[t!]
\centering
\hspace{-0.8 cm} \includegraphics[width=0.3 \linewidth]{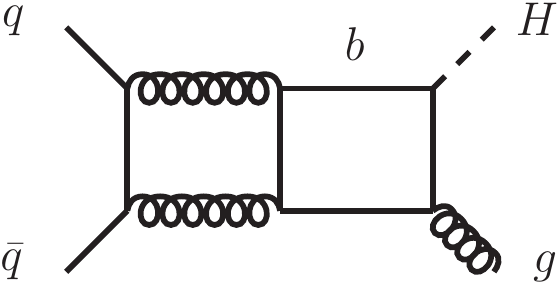} \hspace{0.25 cm}
\includegraphics[width=0.3 \linewidth]{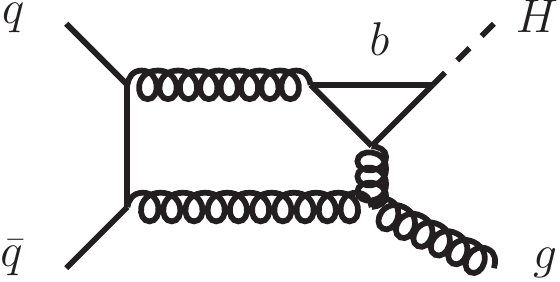} \hspace{0.25 cm}
\includegraphics[width=0.3 \linewidth]{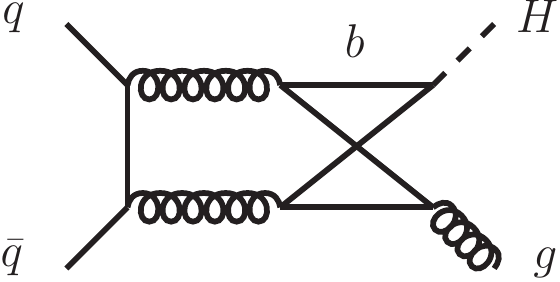}
\caption{Examples of two-loop Feynman diagrams that contribute to the process $H\rightarrow q\bar{q}g$. 
}
\label{fig::feyndiag}
\end{figure}

To compute  the two form factors in Eq.(\ref{eq:bareF}), we proceed as described 
in Ref.~\cite{Melnikov:2016qoc}, where $H \to ggg$  amplitudes were studied.  
The Feynman diagrams that contribute to the process Eq.\eqref{eq:Htoqqg} 
are produced with QGRAF~\cite{Nogueira:1991ex} and  independently with FeynArts~\cite{Hahn:2000kx}. 
Allowing for massless external quarks and both massive and massless internal quarks, we find 
two diagrams at one loop and $49$ at two loops; examples are shown in Figure \ref{fig::feyndiag}. 

The two form factors are extracted by applying projection 
operators to  individual Feynman diagrams. 
We use the same notation as in Ref.~\cite{Gehrmann:2011aa}
and define
\begin{equation}
T_1^\mu = \bar{u}(p_1) \tau_1^\mu v(p_2)\,, \qquad T_2^\mu = \bar{u}(p_1) \tau_2^\mu v(p_2)\,.
\end{equation}
In terms of $T_{1,2}^\mu$,  the projection operators read
\begin{eqnarray}
\mathcal{P}^\mu(F_1) &=& \frac{1}{2(d-3)st} \left[ \frac{(d-2)}{t}(T_1^\mu)^{\dagger}
-\frac{(d-4)}{u}(T_2^\mu)^{\dagger} \right], \\
\mathcal{P}^\mu(F_2) &=& 
\frac{1}{2(d-3)su} \left[ \frac{(d-2)}{u}(T_2^\mu)^{\dagger}-\frac{(d-4)}{t}(T_1^\mu)^{\dagger} \right].
\end{eqnarray}
Their action on the amplitude is described  by the following formula 
\be
F_i(s,t,u,m_b)=\sum_{pol}\mathcal{P}^\mu(F_i) 
(\epsilon_{3,\mu}(p_3))^* \epsilon_3^\nu(p_3) \, \mathcal{A_\nu}(s,t,u,m_b)\,,
\ee
where sums over quark, antiquark and gluon polarizations need to be computed. 
These  polarization sums are calculated  with the help of standard formulas 
\begin{gather}
\sum_{pol} u(p_1)\bar{u}(p_1) = \p_1, \quad \sum_{pol} v(p_2)\bar{v}(p_2) = \p_2 \,, \label{eq:polsums2} \\
\sum_{pol} \left( \epsilon_3^{\mu}(p_3) \right)^* \epsilon_3^{\nu}(p_3) = 
- g^{\mu \nu}\,.
\label{eq:polsums3}
\end{gather}
We note that it is  allowed to use unphysical result for the sum over gluon polarizations 
as in Eq.\eqref{eq:polsums3}  since 
the tensor structures $T_{1,2} ^\mu$ satisfy the transversality condition independently.

The algebraic manipulations required to apply the projection operators to the amplitudes,
perform the polarization sums and extract the form factors have been carried out  independently using  
 both FORM~\cite{Vermaseren:2000nd} and FormCalc~\cite{Hahn:1998yk}. 
After performing the Lorentz algebra, the form factors are expressed as linear combinations of scalar integrals
\begin{equation}
\I_{\rm top}(a_1,a_2,...,a_8,a_9)
= \int 
\frac{\mathfrak{D}^dk \mathfrak{D}^dl}{[1]^{a_1} [2]^{a_2}  [3]^{a_3}  [4]^{a_4}  [5]^{a_5}  [6]^{a_6}  [7]^{a_7}  [8]^{a_8}  [9]^{a_9} 
},
\label{eq3.5a}
\end{equation}
with the integration measure defined as 
\be
\mathfrak{D}^dk = (-m_h^2)^{(4-d)/2}\frac{(4\pi)^{d/2}}{i\Gamma(1+\epsilon)} \int \frac{d^dk}{(2\pi)^d}.
\ee

\begin{table}[th!]
\begin{center}
\begin{tabular}{| c | l | l | l|}
\hline
Prop. & Topology PL1 & Topology PL2 & Topology NPL \\
\hline
$[1]$ & $ k^2$ &                                  $k^2-m_b^2$  &                         $k^2-m_b^2$ \\
$[2]$ & $(k-p_1)^2$ &                         $(k-p_1)^2-m_b^2$  &               $(k+p_1)^2-m_b^2$ \\
$[3]$ & $(k-p_1-p_2)^2$ &                   $(k-p_1-p_2)^2-m_b^2$  &        $(k-p_2-p_3)^2-m_b^2$ \\
$[4]$ & $(k-p_1-p_2-p_3)^2$ &           $(k-p_1-p_2-p_3)^2-m_b^2$  & $l^2-m_b^2$ \\
$[5]$ & $l^2-m_b^2$ &                         $l^2-m_b^2$  &                         $(l+p_1)^2-m_b^2$ \\
$[6]$ & $(l-p_1)^2-m_b^2$ &                $(l-p_1)^2-m_b^2$  &                $(l-p_3)^2-m_b^2$ \\
$[7]$ & $(l-p_1-p_2)^2-m_b^2$ &         $(l-p_1-p_2)^2-m_b^2$  &        $(k-l)^2$ \\
$[8]$ & $(l-p_1-p_2-p_3)^2-m_b^2$ &  $(l-p_1-p_2-p_3)^2-m_b^2$  & $(k-l-p_2)^2$ \\
$[9]$ & $(k-l)^2-m_b^2$ &                     $(k-l)^2$  &                               $(k-l-p_2-p_3)^2$ \\
\hline 
\end{tabular}
\caption{Feynman propagators of the three integral families, see Eq.\eqref{eq3.5a}.} \label{tab:topos}
\end{center}
\end{table}

\noindent
As discussed in Ref.~\cite{Melnikov:2016qoc},
the  scalar integrals shown in Eq.\eqref{eq3.5a} can be organized into three integral families. 
For convenience we provide their definitions in Table \ref{tab:topos}.

Similarly to the case of   the $H \to ggg$ amplitude, 
all scalar integrals that are required for the $H \to q \bar q g$ amplitude 
can be reduced to master integrals (MIs) 
using  integration by parts identities~\cite{Tkachov:1981wb,Chetyrkin:1981qh}.
However,  in the  $qg$ case,  the most complicated non-planar sector does not contribute
to the amplitude and the reduction can be performed  entirely using the public codes 
FIRE5~\cite{Smirnov:2008iw,Smirnov:2014hma} and 
Reduze2~\cite{Bauer:2000cp,Studerus:2009ye,vonManteuffel:2012np,fermat}.
All the relevant MIs have been computed in~Ref.\cite{Melnikov:2016qoc} 
with the method of differential equations,  as an expansion in  $\kappa$.
As a result we obtain analytic expressions for the form factors to leading order in $\kappa$
and up to order $\epsilon^2$ and $\epsilon^0$ for their one- and two-loop contributions,  respectively.
It is well known, that the expansion for small values of $\kappa$ is non-analytic and the one- and
two-loop form factors develop logarithmic singularities $\propto \log{(\kappa)}$ as $\kappa \to 0$.
Written in terms of $m_h^2$ and the dimensionless variables $y,z,\kappa$,
the unrenormalized form factors have the following expansion
\begin{align}
\lim_{m_b  \to 0} F_j^{(1),\text{un}}(y,z,\kappa,m_h^2) &= \frac{m_b^2}{v}\, 
\frac{1}{m_h^4} \sum_{n=0}^2\; \epsilon^n\; \sum_{a=0}^4\, f_{a,j}^{(1l,n)}(y,z)\, \log^a{\kappa}\,, \label{eq:expform1} \\
\lim_{m_b  \to 0} F_j^{(2),\text{un}}(y,z,\kappa,m_h^2) &= \frac{m_b^2}{v} \,
\frac{1}{m_h^4} \sum_{n=-2}^0\; \epsilon^n\; \sum_{a=0}^4\, f_{a,j}^{(2l,n)}(y,z)\, \log^a{\kappa}, \label{eq:expform2}
\end{align}
where $v$ is the Higgs vacuum expectation value. One of the two powers of the bottom quark mass $m_b$ 
shown in Eqs.(\ref{eq:expform1},~\ref{eq:expform2})  has its  origin 
in the  Yukawa coupling $b \bar b H $ and the  other in the helicity flip  on one of the bottom quark lines 
required to enable the $ggH$ coupling through the bottom quark loop.
The coefficients in this expansion $f^{(1l,n)}_{a,j}(y,z)$ and $f^{(2l,n)}_{a,j}(y,z)$ 
can all be expressed in terms of a subset of Goncharov polylogarithms known as 
2dHPLs. They were defined in Ref.~\cite{Gehrmann:2000zt}.
We refer 
to Ref.~\cite{Melnikov:2016qoc} for additional  details on how  to efficiently compute 
MIs in the limit of vanishing quark masses,  $\kappa \to 0$.

\section{Ultraviolet renormalization  and extraction of infrared singularities}
\label{sec:UVIR} \setcounter{equation}{0} 
\numberwithin{equation}{section} 

The bare form factors are regularized dimensionally and contain poles in $\epsilon=(4-d)/2$. 
These poles 
originate from ultraviolet (UV) and infrared (IR) divergences 
of loop integrals that contribute to the scattering amplitude.  These divergences are either 
removed by UV renormalization or cancel against real emission contributions when physical 
cross sections are calculated. The relevant information is contained in suitably-defined 
finite parts of the amplitudes whose computation we now describe. 

To render the amplitudes finite,  
we first subtract the UV poles and write the UV renormalized form factors as
\begin{equation}
F_j^{\uv}(s,t,u,m_b) = \sqrt{\frac{ \alpha_s^3}{\pi\, S_\epsilon^3}}\,  
\left[  F_j^{(1),\uv}
+  \left( \frac{\alpha_s}{2 \pi}  \right) F_j^{(2),\uv}  + \mathcal{O}(\alpha_s^3) \right]\,.
\end{equation}
Renormalized form factors are obtained from the bare ones 
 $F_j^{\un}$ in Eq.\eqref{eq:bareF} by expressing  bare parameters in 
terms of their renormalized counterparts and including the  wave-function 
renormalization factor for each external gluon. 
The massless quark contributions to the coupling constant are renormalized in the $\overline{\text{MS}}$-scheme, 
while the bottom-quark contribution is  renormalized at 
zero-momentum transfer. 
Choosing a  ``physical'' renormalization scheme for  the bottom quark mass is more tricky, 
see discussion in Ref.~\cite{Melnikov:2016qoc}.
For simplicity, we renormalize the bottom mass in the on-shell scheme.
Relations between  bare and renormalized 
parameters are described by the following equations 
\begin{gather}
\alpha_0\, \mu_0^{2 \epsilon} \; S_\epsilon = \alpha_s\, \mu_R^{2 \epsilon}\, 
\left[ 1 - \frac{1}{\epsilon} \left( \beta_0 
+ \delta_w  \right ) \; \left( \frac{\alpha_s}{2 \pi}  \right) + \mathcal{O}(\alpha_s^2)\right], \\
m_{b,0} = m_b\left[ 1 +  \left( \frac{\alpha_s}{2 \pi}  \right)\, \delta_m + \mathcal{O}(\alpha_s^2)\; \right]\,,
\label{eq:baren}
\end{gather}
where $S_\epsilon = (4 \pi)^\epsilon\, e^{-\epsilon \, \gamma_E}\,, \;\; \gamma_E = 0.5772..$,
$\beta_0 = 11/6\; C_A - 2/3\, T_R \, N_f$, $T_R = 1/2$. In addition,  $C_A=N_c$ is the number of 
colors and $N_f$ is the number of massless quark flavors. The renormalization constants read 
\be
\delta_w =  -2/3\;T_R (m_b^2/\mu_R^2)^{-\epsilon}, \quad \delta_m = C_F
\left( \frac{m_b^2}{\mu_R^2}\right)^{-\epsilon} 
 \left(  -\frac{3}{2\epsilon} -2 + \mathcal{O}(\epsilon) \right).\, 
\ee
Gluon wave function renormalization is performed by multiplying the 
form factors with 
$$Z_A^{1/2} = 
\left( 1 + \left( \frac{\alpha_s}{2 \pi}  \right)\,
\delta_w  + \mathcal{O}(\alpha_s^2)\right)^{1/2}
= 1 + \frac{1}{2} 
\left( \frac{\alpha_s}{2 \pi}  \right)\, \delta_w + \mathcal{O}(\alpha_s^2)\,, $$
for each external gluon in the process.
Putting everything together,  we find that the UV-renormalized and bare form factors are related by 
\begin{align}
 F_j^{(1),\uv} &=  F_j^{(1),\un}\,, \nonumber \\
 F_j^{(2),\uv} &= S_\epsilon^{-1}\, F_j^{(2),\un}  
 - \left(\frac{3 \,\beta_0}{2\,\epsilon} + \frac{\delta_w}{\epsilon}\right)\; F_j^{(1),\un} 
 + \, m_b\; \frac{dF_j^{(1),\un}}{d m_b}\, \delta_m  \label{eq:UVren}
 \,.
\end{align}

Even after UV renormalization, the form factors still contain $1/\epsilon$ divergences 
that reflect the infrared singularities of scattering amplitudes. These infrared 
singularities are universal~\cite{Catani:1998bh};  this implies that 
infrared divergences of a
two-loop QCD amplitude for a given process can be predicted in terms of 
 Born  and one-loop amplitudes for  that process.  
Since the amplitude $0 \to H+q \bar q +g$ appears at one-loop, 
the infrared structure of the two-loop amplitude is that of next-to-leading order 
and, therefore, simple.  We have~\cite{Catani:1998bh}
\be
F_j^{(1), \uv} = F_j^{(1), \fin} \,,\;\;\;\;\;
F_j^{(2),\uv} = I_1(\epsilon)  F_j^{(1),\uv} + F_j^{(2), \fin}  \, \label{eq:IRsub},
\ee
where $F_{1,2}^{\rm fin}$ are finite in the limit $\epsilon \to 0$. 
The  operator $I_1(\epsilon)$ contains all infrared singularities; for 
 a process with a quark, antiquark and a gluon it assumes the following form 
\begin{eqnarray}
I_1(\epsilon) = -\frac{ e^{\epsilon \gamma}}{2 \Gamma(1-\epsilon)} && \hspace{-0.5cm}
\left( C_A\left( \frac{1}{\epsilon^2} + \frac{3}{4\,\epsilon} + \frac{\beta_0}{2\,C_A\,\epsilon}\right)\; \left(\left(-\frac{t}{\mu_R^2}\right)^{-\epsilon} + \left(-\frac{u}{\mu_R^2}\right)^{-\epsilon}\right) \right. \nonumber\\
& & \hspace{-0.3cm} \left. -\frac{1}{C_A}\left( \frac{1}{\epsilon^2} + \frac{3}{2\,\epsilon}\right)\; 
\left(-\frac{s}{\mu_R^2}\right)^{-\epsilon} \right) .
\label{eq:cataniI1}
\end{eqnarray}
We note that because of the $\epsilon^2$ poles that 
appear in the $I_1(\epsilon)$ operator, we require the 
one-loop amplitude to order
$\epsilon^2$,  as indicated in Eq.\eqref{eq:expform1}. 

\section{Helicity amplitudes}
\label{sec:helamp} \setcounter{equation}{0} 
\numberwithin{equation}{section}

For phenomenological applications, 
it is useful to derive analytic  expressions for the helicity amplitudes
for the process $H \to q \bar{q} g$. We begin by expressing the helicity amplitudes  
through the form factors defined in Eq.\eqref{eq:tensor}.
We use the spinor-helicity formalism (see e.g.~\cite{Dixon:1996wi}) and define positive and negative helicity
spinors for  massless external quarks as
\begin{align}
u_{+}(p) &= v_{-}(p) = | p \rangle \,, \quad u_{-}(p) = v_{+}(p) = | p ]\,,\nonumber \\
\bar{u}_{+}(p) &= \bar{v}_{-}(p) = [ p | \,, \quad \bar{u}_{-}(p) = \bar{v}_{+}(p) = \langle p |\,.
\label{eq:polquark}
\end{align}
For the external gluon we define
\begin{equation}
\epsilon_{3,+}^{\mu}(p_3) = \frac{\langle q | \gamma^\mu | 3 ]}{\sqrt{2} \langle q \, 3 \rangle }
\,, \qquad
\epsilon_{3,-}^{\mu}(p_3) = - \frac{[ q | \gamma^\mu | 3 \rangle}{\sqrt{2} [ q \, 3 ] }\,,
\label{eq:polglu}
\end{equation}
where $q$  is an arbitrary (light-like) reference vector.

We define the helicity amplitudes as 
\begin{equation}
\mathcal{A}_{\lambda_1 \lambda_2 \lambda_3}(s,t,u,m_b) = 
\epsilon_{3,\lambda_3}^\mu (p_3)  
\bar{u}_{\lambda_1}(p_1)\, \mathcal{A}_\mu(s,t,u,m_b)\, v_{\lambda_2}(p_2)\,.
\label{eq:hela}
\end{equation}
Since QCD interactions do not change the helicities of massless fermions, 
the helicities of the quark and the antiquark in Eq.\eqref{eq:hela} are correlated. This implies that 
there are, in total, only four possible helicity configurations. 
Out of these four, only one is independent; the other 
three can be obtained from it by charge and parity conjugation. 
We choose the amplitude $\mathcal{A}_{- + +}$ as an independent 
and obtain 
\begin{align}
\mathcal{A}_{-++}(s,t,u,m_b) = \frac{1}{\sqrt{2}} \frac{[23]^2}{[12]\, m_h^2}\, \Omega_{-++}(s,t,u,m_b), 
\label{eq:helampl}
\end{align}
where the helicity coefficient $\Omega_{-++}(s,t,u,m_b)$ is dimensionless. 
The amplitudes for the other helicity assignments can be obtained from
$\mathcal{A}_{- + +}$ by complex conjugation and permutation of the external legs as follows
\begin{align}
\mathcal{A}_{+-+}(p_1,p_2,p_3) & = \mathcal{A}_{-++}(p_2,p_1,p_3)\,, \\
\mathcal{A}_{+--}(p_1,p_2,p_3) & = \left[ \mathcal{A}_{-++}(p_1,p_2,p_3)\right]^*\,, \\
\mathcal{A}_{-+-}(p_1,p_2,p_3) & = \left[ \mathcal{A}_{-++}(p_2,p_1,p_3)\right]^*\,.
\label{eq:allhelampl}
\end{align}
Note that complex conjugation must be performed only on the spinor-helicity structures and not
on the helicity coefficient $\Omega_{-++}$.
We express 
the helicity coefficient in terms of form factors and find 
\begin{equation}
\Omega_{-++} = s\,m_h^2\, F_1\,.
\end{equation}

When expanding the UV-renormalized helicity coefficient $\Omega_{-++}$
in the strong coupling constant, 
it is convenient to  factor out the overall coefficient $m_b^2/v$. We obtain 
\begin{align}
\Omega_{-++} = \frac{m_b^2}{v} \, \sqrt{\frac{\alpha_s^3}{\pi}} \left[ \Omega_{-++}^{(1l)}
+ \frac{\alpha_s}{2 \pi} \Omega_{-++}^{(2l)} + \mathcal{O}(\alpha_s^2)\right]\,.
\end{align}

The ultraviolet renormalization and subtraction 
of infrared singularities described in the context of  form factors, 
can be  applied verbatim 
to the helicity coefficient $\Omega$. Following the discussion in the previous 
Section, we write  
\begin{align}
\Omega_{-++}^{(2l)} = I_1(\epsilon) \Omega_{-++}^{(1l)} + \Omega_{-++}^{(2l),\fin},
\end{align}
where the operator $I_1(\epsilon)$ is defined in Eq.\eqref{eq:cataniI1}.

We renormalize the coupling constant at the scale $\mu = m_h$ in a theory 
with $N_f$ active flavors.
Since we are interested in the kinematic  region where 
all scales are much larger than the bottom quark mass $m_b$, it is reasonable  to perform
a scheme change and define the amplitude in terms of the strong coupling constant which
evolves with $N_f + 1$ active flavors.
At the scale $\mu = m_h$, the relation is very simple and reads
\begin{equation}
\alpha_s^{(N_f)} = \alpha_s^{(N_f+1)} \left[ 1 - \frac{\alpha_s^{(N_f+1)}}{6\, \pi}
\log{\left( \frac{m_h^2}{m_b^2}\right)} + \mathcal{O}(\alpha_s^2)\right] \label{eq:NfNfp1}.
\end{equation}
Eq.\eqref{eq:NfNfp1} implies the following relations for Catani's finite
remainder of the helicity amplitude
\begin{align}
\overline{\Omega}_{-++}^{(1l), \fin} = \Omega_{-++}^{(1l), \fin} \,, \qquad
\overline{\Omega}_{-++}^{(2l), \fin} = \Omega_{-++}^{(2l), \fin} - 
\frac{1}{2} \log{\left( \frac{m_h^2}{m_b^2}\right)}\, \overline{\Omega}_{-++}^{(1l),\fin},
\label{eq:chscheme}
\end{align}
where $\overline{\Omega}$ are the helicity coefficients corresponding to $\alpha_s^{(N_f+1)}$
evolved with $N_f + 1$ active flavors.
We provide Catani's finite remainder of the helicity amplitudes 
defined in the scheme of Eq.\eqref{eq:chscheme} 
together with the arXiv submission of this paper. 

\section{Analytic continuation}
\label{sec:cont}
\numberwithin{equation}{section}

We are interested in computing the two-loop contributions to scattering 
amplitudes for Higgs production processes at the LHC. The calculation of the 
decay amplitudes $H \to q \bar q g$ reported in the previous sections 
allows us to compute the scattering amplitudes for the three partonic 
processes  $q \bar q \to H g$, $q g \to H q$ and $\bar q g \to H  \bar q$,  
by crossing different particles from final to initial states and performing the relevant analytic 
continuation. 

To describe the analytic continuation, 
we follow the notation introduced in Ref.~\cite{Gehrmann:2002zr}. 
To account for all helicity amplitudes of the two scattering processes 
$q \bar{q} \to H g$ and $q g \to H q$,
we need to consider three kinematic  regions that 
we will refer to as  $(2a)_+$, $(3a)_+$ and $(4a)_+$, while the kinematic region of 
the decay process is referred to as $(1a)_+$.
The regions are defined as 
\begin{align}
\mbox{region} (1a)_+ &: \qquad H(p_4) \to q(p_1) + \bar{q}(p_2) + g(p_3), \\
\mbox{region} (2a)_+ &: \qquad q(p_2) + \bar{q}(p_1) \to H(p_4) + g(p_3),\\
\mbox{region} (3a)_+ &: \qquad q(p_1) + g(p_3) \to H(p_4) + q(p_2), \\
\mbox{region} (4a)_+ &: \qquad q(p_2) + g(p_3) \to H(p_4) + q(p_1).
\end{align}
Note that, in contrast to $gg \to H g$ and $q \bar{q} \to H g$, 
for the  process $q g \to H q$ 
the region $(3a)_+$ is needed 
in order to 
obtain the helicity amplitudes with the opposite helicity assignment for the 
quarks from $\Omega_{-++}$ only. 

The analytic continuation from region $(1a)_+$ to the other  regions is described
in detail in~\cite{Gehrmann:2002zr}. In particular, the spinor 
products in Eq.\eqref{eq:helampl} are unchanged while the Goncharov polylogarithms
appearing in the coefficient $\Omega_{-++}$ develop imaginary parts when computed
in the different regions $(2a)_+$, $(3a)_+$ and $(4a)_+$. The imaginary parts can be
extracted explicitly in terms of real valued functions following the strategy 
outlined in~\cite{Gehrmann:2002zr} using suitable changes of variables. 
Once the imaginary part is extracted, the numerical evaluation of $\Omega_{-++}$ can be performed using
routines presented in~\cite{Gehrmann:2001pz,Gehrmann:2001jv}.

The analytic continuation is best understood by starting with the Euclidean non-physical region 
$(1a)_-$ and defined by 
\be
(1a)_- \;:  \quad m_h^2,s, t,u < 0\,. \\
\ee
In the three ``scattering regions'' 
the Mandelstam invariants become 
\begin{align}
(2a)_+ \; :& \quad m_h^2 >0\,, \quad s >0\,, \quad t,u < 0 \,,\\
(3a)_+ \; :& \quad m_h^2 >0\,, \quad t >0\,, \quad s,u < 0 \,,\\
(4a)_+ \; :& \quad m_h^2 >0\,, \quad u >0\,, \quad s,t < 0\,.
\end{align}
Continuation from region $(1a)_-$ to $(2a)_+$, $(3a)_+$ or $(4a)_+$,
is achieved by providing an infinitesimal positive imaginary part to the following 
invariants
\begin{align}
(2a)_+ \; :& \quad m_h^2 \to m_h^2 + i\, 0 \,,\;\; s \to s + i\, 0 \,,\\
(3a)_+ \; :& \quad m_h^2 \to m_h^2 + i\, 0 \,,\;\; t \to t + i\, 0 \,,\\
(4a)_+ \; :& \quad m_h^2 \to m_h^2 + i\, 0 \,,\;\; u \to u + i\, 0 \,.
\end{align}
For the three ``scattering regions'',  we define  the new variables $u_j$ and $v_j$ as 
\begin{align}
(2a)_+ \; :&\quad u_{2a} = -\frac{t}{s} = - \frac{y}{1-y-z}\,, \qquad v_{2a} = \frac{m_h^2}{s} = \frac{1}{1-y-z}\,,
\label{eq:ancontchv1}
\\
(3a)_+ \; :& \quad u_{3a} = -\frac{u}{t} = - \frac{z}{y}\,, \qquad v_{3a} = \frac{m_h^2}{t} = \frac{1}{y}\,,
\label{eq:ancontchv2}
\\
(4a)_+ \; :& \quad u_{4a} = -\frac{t}{u} = - \frac{y}{z}\,, \qquad v_{4a} = \frac{m_h^2}{u} = \frac{1}{z}\,.
\label{eq:ancontchv3}
\end{align}
These variables satisfy the following constraints 
\be
0 \leq u_j \leq v_j\,, \qquad 0 \leq v_j \leq 1\qquad \mbox{for} \quad j = 2a, 3a, 4a\,.
\ee

In each region the extraction of the imaginary parts in terms of explicitly real-valued
functions is achieved by changing variables in the Goncharov polylogarithms
from $(y,z)$ to $(u_j,v_j)$ with $j = 2a, 3a, 4a$.
As the result, the helicity amplitudes can be written as linear combinations of real-valued Goncharov polylogarithms
of arguments $(u_j,v_j)$.
We provide  the one- and two-loop helicity coefficient $\Omega_{-++}$ in all regions described above 
together with the arXiv submission of this paper.

\section{Collinear limit} 
\label{sec:colimit}

In the kinematic limit of forward scattering the helicity amplitudes simplify. 
To derive the approximation for the amplitude,  we 
start by considering the amplitude $H \to q \bar{q}g$ in the limit where quark and antiquark
are emitted collinearly. By crossing the anti-quark and the gluon to the initial state, 
we then obtain the scattering amplitude for the partonic process 
$q g \to H q $ in the collinear approximation. 

More precisely, for the decay amplitude,  we are interested in the
situation $m_b^2 \ll s \ll m_h^2  \sim t \sim u$, which implies $1-y-z = x \to 0$ 
and $z \sim y =(1-z) \sim \mathcal{O}(1)$.
For simplicity we define the abbreviations
\be
L = \log{\left( \kappa \right)} = \log{ \left( \frac{-m_b^2}{m_h^2}\right)  }\, 
\qquad \mbox{and} \qquad 
\eta = \frac{ \log{\left( x/ \kappa \right) }}{\log \left( \kappa \right) }\,.
\ee
In the collinear  limit $L \gg 1$ while $\eta \sim 1$. Expanding the helicity amplitude
we find that at one loop all dependence on the other kinematical invariant drops and we are left with
\begin{align}
\overline{\Omega}_{-++}^{(1l),\fin} = L^2 \left( \eta^2 - 1 \right) - 4 \,.
\end{align}
At two loops there is a residual functional dependence on $y$ and the amplitudes 
are still too complicated to be reported here entirely.\footnote{Recall that in the
limit $x \to 0$ we have $y \to 1-z$.} We write therefore the amplitude
keeping the coefficients of the leading, next-to-leading and next-to-next-to-leading
logarithms, which are relatively compact. We find

\begin{align}
\overline{\Omega}_{-++}^{(2l),\fin} &=  i\, \pi \frac{3}{2} \bar{\beta_0} \overline{\Omega}_{-++}^{(1l),\fin} 
+ L^4 \frac{5}{36}  \left( 1 + \eta \right)^2 
\left( 1 - 2 \eta + 3 \eta^2 \right) \nonumber \\
& + L^3 \left( 1 + \eta \right) 
\left[ \left( \eta^2 - 1 \right) \left( \frac{3}{2} \log{(y)} + \frac{3}{2} \log{(1-y)} + \bar{\beta_0} \right)
- \frac{1}{12}\left( 3 + \eta \right) \left( 11 \eta - 15 \right)
\right] 
\nonumber \\
&+ L^2 \left\{ \left( \eta^2 - 1 \right) \left[ 
\frac{29}{24} \left( \log{(y)} + \log{(1-y)} \right) + \frac{3}{2}\left( \Li_2(y) + \Li_2(1-y) \right) \right. \right.
\nonumber \\ & \left. \left.+ \bar{\beta_0} \left( \frac{1}{4}\left( \log{(y)} + \log{(1-y)} \right) - \frac{5}{3} \right)
\right]
+ \frac{14 \pi ^2 \eta ^2}{9}-\frac{7 \eta ^2}{6}+\frac{10 \pi ^2 \eta }{9}+\frac{\pi^2}{9}+\frac{7}{6} \right\}\,.
\label{eq:coll_lim}
\end{align}

To obtain the scattering amplitude for the physical scattering process $q g \to q  H$ 
(or similarly for $\bar{q} g \to \bar{q} H$) in the 
limit of a small momentum transfer from the quark line to the gluon-Higgs line one can of course use
the general change of variables defined in Eqs.(\ref{eq:ancontchv2},~\ref{eq:ancontchv3}).
Nevertheless, in this limit we have to deal with much simpler functions of one variable only
(in general, harmonic polylogarithms) 
and it is simpler to replace $y \to -\widetilde{y} + i 0$ and $m_h^2 \to m_h^2 + i 0$ directly in Eq.\eqref{eq:coll_lim}. 
The sign of the imaginary part to be associated to $y$ is determined noticing that in the 
scattering region
$m_h^2 > 0$ and $t = - \widetilde{t} < 0$ and therefore
\be
y \to -\frac{(-t)}{m_h^2 + i 0} = -\frac{\widetilde{t}}{m_h^2} + i\, 0 = -\widetilde{y} + i\, 0.
\ee

\section{Conclusions}
\label{sec:conclusions} \setcounter{equation}{0} 
\numberwithin{equation}{section} 
We described the calculation of the two-loop contribution to the scattering amplitude 
of the process   $H \to q \bar{q} g$ mediated by a massive quark loop,
in the kinematic limit where the quark mass is the smallest parameter in the process.
For a typical hard 
LHC collision, this kinematic  limit is expected to describe very well
the contribution of loops of bottom quarks.
The results  presented in this paper  provide the last missing ingredient required to obtain  
all the two-loop scattering amplitudes for the NLO QCD 
corrections to the top-bottom interference in Higgs plus jet production at the LHC.

In addition to their potential phenomenological relevance, these  results are also interesting 
as a step towards understanding the structure of  large logarithmic corrections
that appear in amplitudes with virtual quarks loops that require a helicity flip. 
Resummation of these logarithms is not understood, and the results presented
in this paper may help in further studies  of this problem. 

\section*{Acknowledgements}

We would like to thank T. Hahn for his help with FormCalc. The research of K.M. was supported by 
the German Federal Ministry for Education $\&$ Research (BMBF) under grant O5H15VKCCA.

\appendix

\bibliographystyle{bibliostyle}   
\bibliography{Biblio}
\end{document}